# Effects of Crystalline Disorder on Interfacial and Magnetic Properties of Sputtered Topological Insulator/Ferromagnet Heterostructures

[1]Nirjhar Bhattacharjee, [3]Krishnamurthy Mahalingam, [2]Adrian Fedorko, [1]Alexandria Will-Cole, [1]Jaehyeon Ryu, [3]Michael Page, [3]Michael McConney, [1†]Hui Fang, [2]Don Heiman, [1]Nian Xiang Sun*

1 Northeastern University, Department of Electrical and Computer Engineering, Boston MA 02115

2 Northeastern University, Department of Physics, Boston MA 02115

3 Air Force Research Laboratory, Nano-electronic Materials Branch, Wright Patterson Air Force Base, OH 05433

ABSTRACT. Thin films of Topological insulators (TIs) coupled with ferromagnets (FMs) are excellent candidates for energy-efficient spintronics devices. Here, the effect of crystalline structural disorder of TI on interfacial and magnetic properties of sputter-deposited TI/FM, $Bi_2Te_3/Ni_{80}Fe_{20}$, heterostructures is reported. Ni and a smaller amount of Fe from Py was found to diffuse across the interface and react with $Bi_2Te_3$. For highly crystalline *c*-axis oriented $Bi_2Te_3$ films, a giant enhancement in Gilbert damping is observed, accompanied by an effective out-of-plane magnetic anisotropy and enhanced damping-like spin-orbit torque (DL-SOT), possibly due to the topological surface states (TSS) of $Bi_2Te_3$. Furthermore, a spontaneous exchange bias is observed in hysteresis loop measurements at low temperatures. This is because of an antiferromagnetic topological interfacial layer formed by reaction of the diffused Ni with $Bi_2Te_3$ which couples with the FM, $Ni_{80}Fe_{20}$. For increasing disorder of $Bi_2Te_3$, a significant weakening of exchange interaction in the AFM interfacial layer is found. These experimental results Abstract length is one paragraph.

## 1. INTRODUCTION

Topological insulators (TIs) of the $(Bi,Sb)_2(Te,Se)_3$ family of compounds are van der Waals (vdW) chalcogenide materials with tetradymite structures. TIs possess large spin-orbit coupling (SOC) resulting in dissipationless surface conducting states – topological surface states (TSS) [1-3]. Introducing magnetic order in TIs leads to gap opening in the TSS bands and possibility of dissipationless quantum anomalous Hall (QAH) and axion insulator states [4-21]. Stimulated by these remarkable material properties, TIs are regarded as promising candidates for realization of energy efficient spintronic devices. TIs possess highly reactive surfaces, thus making them susceptible to formation of interfacial phases when coupled with metallic films [22-25]. Because of their composition, these interfacial layers have the potential for hosting fascinating topological magnetic phases [25]. The majority of reported experiments have studied TIs grown from Molecular Beam Epitaxy (MBE) [22-24], which is a standard technique for growing high-quality, crystalline-ordered thin films. However, MBE suffers from low throughput and is constrained by sample dimensions, making it incompatible for integration in industrial CMOS processes. Magnetron sputtering on the other hand is the semiconductor industry's accepted thin film deposition technique because of its advantage of high throughput and large area film growth. Sputtering also allows easy deposition of TIs with varying crystalline disorder [25-29]. This controllability opens up the possibility of exploration of their disorder-dependent electronic and magnetic properties.

Recently, the topological antiferromagnetic (AFM) compound $NiBi_2Te_4$ was discovered in the interface of highly *c*-axis-oriented sputtered $Bi_2Te_3$/$Ni_{80}Fe_{20}$ heterostructures [25]. Ni from the $Ni_{80}Fe_{20}$ (Py) layer diffuses and reacts with $Bi_2Te_3$ layer, and the reaction is promoted by the delocalized TSS electrons [23-25]. Also, recent experiments have shown the presence of TSS even in amorphous $Bi_2Se_3$ [30]. In this work, the effects of crystalline structural disorder on the interface and magnetic properties of $Bi_2Te_3$/Py heterostructures are investigated. The magnetic species, largely Ni and smaller amounts of Fe, are found to diffuse across the interface into $Bi_2Te_3$, resulting in a magnetic interfacial layer. For increasing *c*-axis-oriented texture of $Bi_2Te_3$, increasing amounts of diffused magnetic species were found to react with $Bi_2Te_3$, which also leads to enhanced magnetic properties. This phenomenon was identified in room temperature hysteresis loop measurements of the magnetic moment versus applied magnetic field, $m(H)$, for the $Bi_2Te_3$/Py samples compared to a Py control sample. As a result of the diffusion of the magnetic species (Ni, Fe) and reaction with $Bi_2Te_3$, the saturation magnetic moment ($m$) is reduced by $\Delta m$ in the $Bi_2Te_3$/Py compared to Py samples suggesting change in valence state of the magnetic species. The values of $\Delta m$ becomes smaller for increasing disorder in $Bi_2Te_3$ suggesting lesser reaction between diffused Ni, Fe and $Bi_2Te_3$. Further, a giant enhancement in Gilbert damping, an out-of-plane canting of magnetization and enhanced DL-SOT were observed in samples with highly *c*- axis oriented TI. However, with significantly reduced crystallinity, surprisingly the granular $Bi_2Te_3$ samples had a comparable enhanced spin-charge conversion efficiency as samples with highly *c*-axis-oriented $Bi_2Te_3$, possibly due to the quantum confinement effect in smaller crystallite grains [26,27]. Low-temperature $m(H)$ and $m(T)$ measurements revealed an AFM ordered phase in the predominantly Ni-diffused $Bi_2Te_3$ interface from the formation of the topological AFM compound $NiBi_2Te_4$ [25]. Interestingly, the strength of the exchange interaction of the interfacial AFM phase, as monitored by the exchange bias, was found to weaken significantly with increase in disorder of the $Bi_2Te_3$ layer. These results indicate strong topological property of TIs with high crystalline *c*-axis-oriented growth, which weakens considerably with increasing crystalline disorder. These experimental results show the possibility of tailoring topological properties of TIs by control of crystalline structural disorder.

## 2. EXPERIMENTAL RESULTS AND DISCUSSIONS

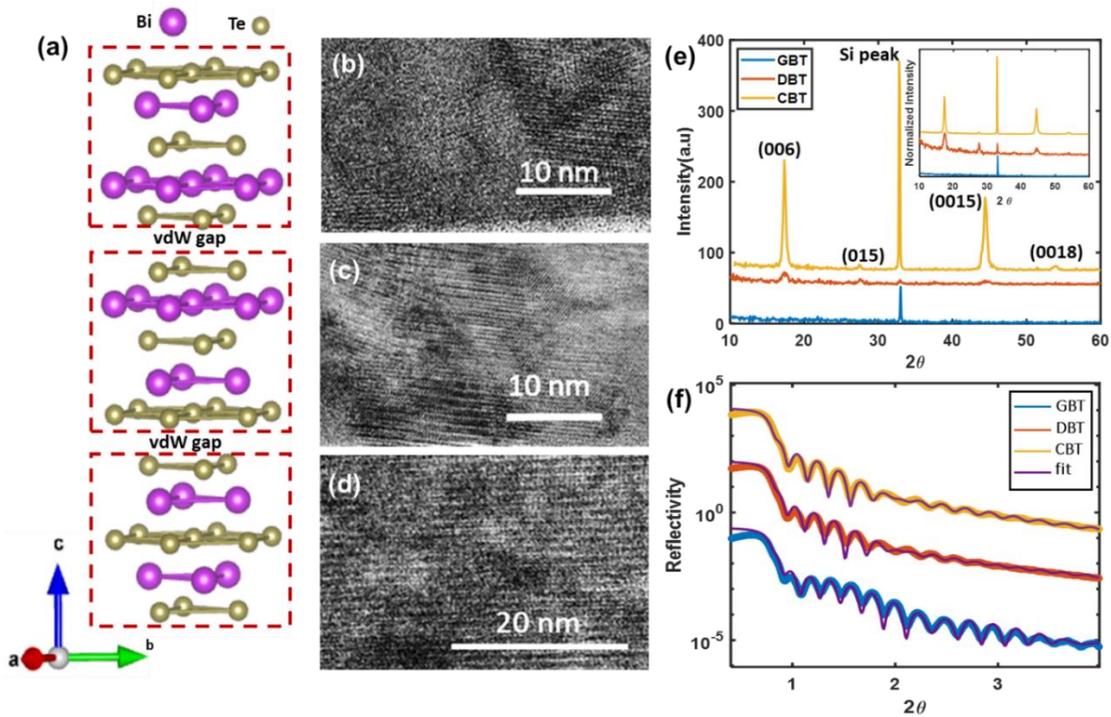

***Figure 1. a)*** *Schematic model of three quintuple $Bi_2Te_3$ unit cells. Cross-sectional HRTEM images showing structural disorder in **b)** GBT, **c)** DBT and **d)** CBT samples. **e)** XRD data for the GBT, DBT and CBT samples. Inset: normalized plots of symmetric XRD data. **f)** XRR plots and theoretical fitting for GBT, DBT and CBT samples used for characterization of thickness and surface roughness. The data for CBT samples are similar to the ones in ref [25].*

**2.1. Crystalline Structure Properties of Sputter-deposited $Bi_2Te_3$.** Samples of 30 nm $Bi_2Te_3$ with varying crystalline disorder, (1) *granular* (GBT), (2) randomly oriented polycrystalline *disordered* (DBT), and (3) highly *c*-axis-oriented *crystalline* (CBT) were grown using RF magnetron sputtering on amorphous thermally oxidized Si substrates (see Supporting Information Section S1 for grain size characterization). Crystalline structural property of the $Bi_2Te_3$ samples were verified using X-ray diffraction (XRD) and high-resolution transmission electron microscope (HRTEM) imaging measurements, as shown in Fig. 1a,b,d. The GBT samples did not show any significant diffraction peaks in the XRD measurement, suggesting a high amorphous content. Further, HRTEM images of the GBT, DBT and CBT samples shown in Fig. 1a verifies the granular, randomly oriented vdW domains and high *c*-axis-oriented layered structure, respectively. The thickness and surface roughness of the samples were characterized using X-ray reflectometry (XRR) measurements, as shown in Fig. 1e. From the fitting of XRR data, thickness of ~30 nm was obtained for all three samples. The fits to the XRR data also revealed surface roughness of 0.7 nm, 1.7 nm and 1.0 nm for the GBT, DBT and CBT samples, respectively, which are typical surface roughness values for sputter-deposited thin films, confirming growth of high-quality TI films.

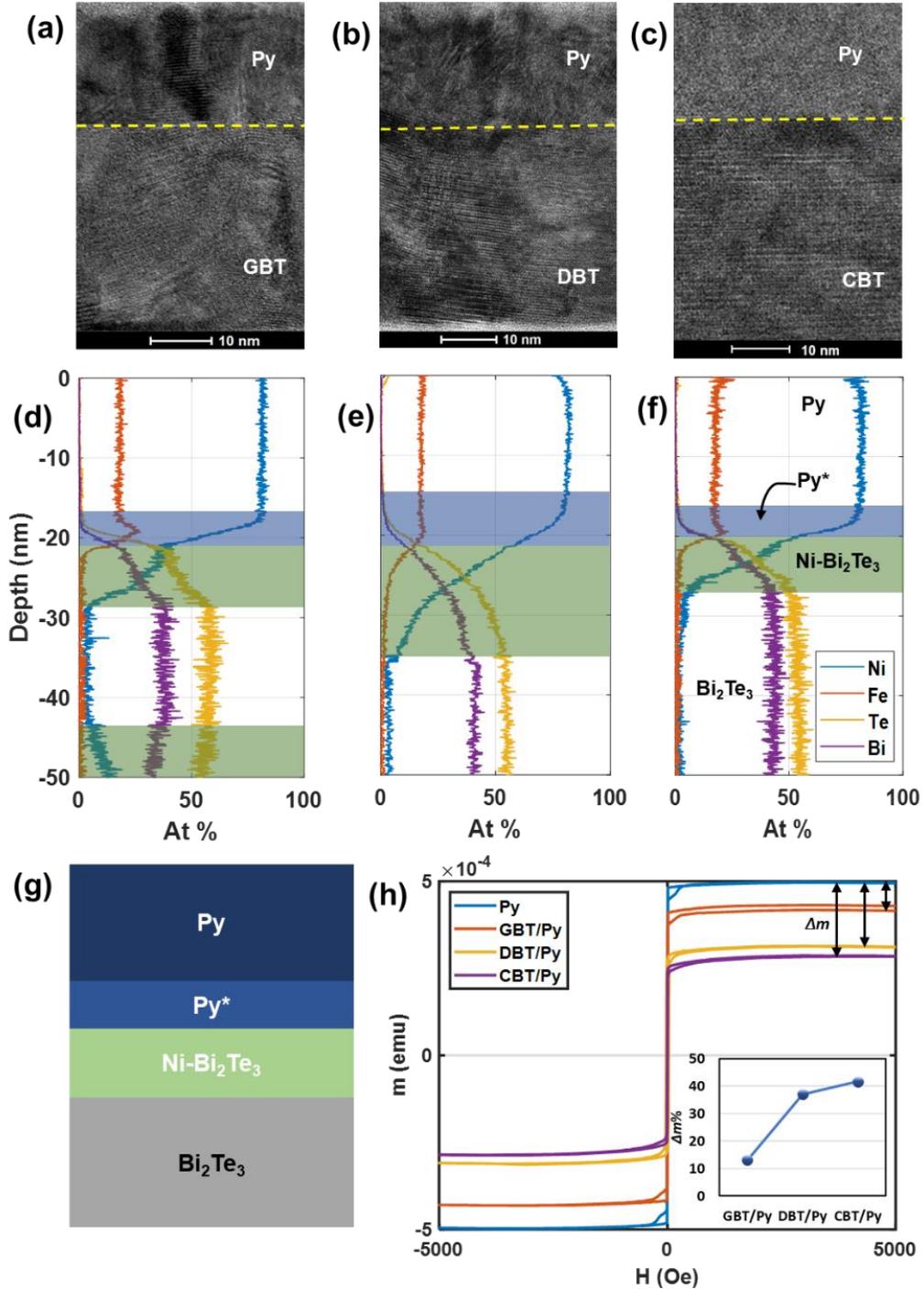

*Figure 2.* Cross-sectional HRTEM images of **a)** highly amorphous GBT/Py, **b)** disordered DBT/Py with randomly oriented vdW domains and **c)** highly c-axis oriented CBT/Py (similar to ref [25]). The yellow dashed lines mark the approximate interface between the $Bi_2Te_3$ and Py layers. Atomic % (At %) characterized using EDS for **d)** GBT/Py, **e)** DBT/Py and **f)** CBT/Py samples. The

*interface layers are highlighted in green and blue colors for the Ni-Bi$_2$Te$_3$ and Py\*, respectively.* ***g)*** *Schematic of the Bi$_2$Te$_3$/Py samples with the layers marked.* ***h)*** *m(H) measurements for IP orientation at room temperature showing loss of moments in the Bi$_2$Te$_3$/Py samples due to interfacial Ni and Fe diffusion and reaction with Bi$_2$Te$_3$. Inset: comparison of % loss of moments (Δm%) in GBT/Py, DBT/Py and CBT/Py samples compared to the control Py sample. The data presented for the CBT sample are similar to the ones in ref [25].*

**2.2. Morphology of Interfacial Layer formed by Ni Diffusion into Bi$_2$Te$_3$.** Heterostructure samples of GBT/Py, DBT/Py and CBT/Py were grown where the thickness of the layers was maintained at 30 nm and 20 nm respectively for Bi$_2$Te$_3$ and Py respectively. HRTEM imaging and energy dispersive X-ray spectroscopy (EDS) measurements were performed to characterize the morphology and stoichiometric composition along the cross section of the samples, as shown in Figures 2a-f (see Supporting Information Table S1 for average atomic %). The HRTEM images in Figures 2 a-c clearly show a highly amorphous nature of GBT, randomly oriented vdW layered crystalline domains in DBT and highly oriented vdW layers in the CBT layers. A closer examination of the interfaces of the heterostructures also reveal a rougher interface in the disordered GBT/Py and DBT/Py samples compared to the CBT/Py sample. The EDS cross-sectional profiles of atomic % of elements in Figures 2d-f show a significant diffusion of Ni (and smaller amounts of Fe) across the TI/FM interface into the Bi$_2$Te$_3$ layers, forming an interfacial layer denoted as Ni-Bi$_2$Te$_3$. In general, the Ni and Fe have a large gradient over a Bi$_2$Te$_3$ distance of 5 to 14 nm, where the Ni averages 30 to 40 %, while the Fe diffusion is much smaller in the GBT/Py and CBT/Py samples. The disordered GBT/Py and DBT/Py samples also have ~3% of Ni diffused throughout the thickness of the Bi$_2$Te$_3$ layer. However, the predominantly Ni-Bi$_2$Te$_3$ layer in the highly ordered CBT/Py sample appears to act as a barrier. This prevents the diffusion of Ni further into the Bi$_2$Te$_3$ bulk. The formation of the Ni-Bi$_2$Te$_3$ layer is likewise accompanied by a thin Fe-rich region in the intermediate Py layer (marked Py*). It is also noted that the moderately-disordered DBT/Py sample which has randomly oriented vdW polycrystalline domains has developed a much higher Ni and Fe diffusion of ~47% and ~9% at the interface, respectively. The Fe diffusion, however, is only ~3-4% at the interface in the GBT/Py and CBT/Py samples.

The diffusion of Ni into high-quality CBT Bi$_2$Te$_3$ was previously shown to result from solid-state reactions leading to the formation of Ni-Te bonds and formation of the topological AFM compound, NiBi$_2$Te$_4$ [25]. Similar to that study, the room temperature $m(H)$ measurements can be used here as an indicator of the reaction of Ni with Bi$_2$Te$_3$ that is promoted by the delocalized TSS electrons. As shown in Figure 2h, all the Bi$_2$Te$_3$/Py samples show a clear decrease in saturation magnetic moment for increasing disorder. This reduction in moments results from change in valence state of the reacting magnetic species. This loss of saturation moment is compared to a control sample of Py by $\Delta m$. The $\Delta m$% values were found to be 13%, 37% and 41% for the GBT, DBT and CBT samples, respectively. This clear enhancement in the loss of moments with crystalline order and hence reactivity of Ni with Bi$_2$Te$_3$ [25] is due to strengthening of TSS with increasing crystallinity of Bi$_2$Te$_3$.

**2.3. Disorder Effects on Room-Temperature Magnetic Properties of Bi$_2$Te$_3$/Py.** To investigate the effects of disorder, $m(H)$ hysteresis loop, and ferromagnetic resonance (FMR) were performed on the three types of samples, highly disordered GBT/Py, moderately disordered DBT/Py and highly ordered CBT/Py. First, $m(H)$ hysteresis loop measurements were performed on the samples, with the magnetic field oriented in-plane (IP) and out-of-plane (OP) relative to the film plane, as shown in Figures 3 a-c. For increasing $c$-axis-oriented growth of the Bi$_2$Te$_3$ layer, the saturation field, $H_s$ measured in the IP and OP configurations show an increasing and decreasing trend, respectively. Also, the ratio of remanence to saturation magnetization ($M_r/M_s$) in IP $m(H)$ loop decreases for increasing crystalline order of $Bi_2Te_3$. These trends indicate an increase in effective OP magnetic easy-axis with increased *c*-axis-oriented texture of $Bi_2Te_3$ in the $Bi_2Te_3$/Py heterostructure samples. This enhanced OP magnetic anisotropy is a characteristic of interaction of the magnetic moments with large SOC in the interfaces [31]. A large OP anisotropy has been previously predicted and observed in other TI/FM-based materials systems [32,33] (also see Supporting Information Section S4). As shown in Figure 3b, the OP $m(H)$ loops for the $Bi_2Te_3$/Py samples also exhibit a smaller hysteresis-loop in the low-field regions. These smaller components of $m(H)$ loop are more prominent in the DBT/Py and CBT/Py samples, which otherwise exhibit a lower OP easy-axis of magnetic moments compared to the highly *c*-axis-oriented CBT/Py sample. The DBT/Py sample also had an unusually large coercive field ($H_c$) resulting from the randomly oriented vdW layered crystalline domains, as observed in both the IP and OP $m(H)$ loop measurements in Figures 3a,b. These effects in the samples with highly disordered TIs are possibly present due to disordered magnetic texture that emerge in their relatively rougher interfaces with a net OP component.

Further information was obtained using ferromagnetic resonance (FMR) measurements to understand the changes in magnetization *dynamics* with changes in TI disorder. The FMR linewidth ($\Delta H$) and resonance field ($H_{res}$) were extracted from the FMR signal at different constant frequencies ($f_{res}$) (Supporting Information Section S3). The Gilbert damping parameter ($\alpha$) was extracted by fitting a straight line to the FMR linewidth versus frequency plot using the equation, $\Delta H = \Delta H_0 + \frac{\gamma}{2\pi}\alpha f_{res}$. Here, $\Delta H_0$ is the inhomogeneous linewidth and $\gamma$ is the gyromagnetic ratio. As shown in Figure 3d, the values of $\alpha$ extracted for the Py, GBT/Py, DBT/Py and CBT/Py samples are 0.0053, 0.0076, 0.0089 and 0.0123, respectively. This shows a progressive increase in

$\alpha$ with increase in crystallite grain size of $Bi_2Te_3$, and a giant enhancement when the $Bi_2Te_3$ layer is highly *c*-axis oriented (summarized in Figure 3f). This effect was also observed in other $Bi_2Te_3$/FM heterostructure materials (Supporting Information Section S5), which signals a large enhancement in SOC and presence of robust TSS in highly *c*-axis-oriented $Bi_2Te_3$. In addition, the effective magnetization, $4\pi M_{eff}$, and the perpendicular magnetic anisotropy (PMA) field, $H_\perp$, were extracted by fitting the modified Kittel equation to the $f_{res}$ versus $H_{res}$ plots shown in Figures 3e,f,

$$f_{res} = \frac{\gamma}{2\pi}\sqrt{(H_{res} + H_a)(H_{res} + H_a + 4\pi M_{eff})},$$ where $4\pi M_{eff} = 4\pi M_s - H_\perp$ and $H_a$ is the uniaxial anisotropy field. The $4\pi M_s$ values were found to be 15.2, 14.7, 14.3 and 14.1 kOe, respectively, for Py, GBT/Py, DBT/Py and CBT/Py samples. The decrease in $4\pi M_s$ for increasing crystalline order demonstrates the reduction in saturation magnetization due to interfacial diffusion of Ni and Fe from Py into $Bi_2Te_3$ [25]. Also, the $H_\perp$ values increased from 4.35, 5.07, 5.54 and 5.79 kOe, for the Py, GBT/Py, DBT/Py and CBT/Py samples, respectively. This enhancement in $H_\perp$ supports the *m*(*H*) results that show an increase in effective OP anisotropy with increasing *c*-axis-oriented texture of $Bi_2Te_3$. The magnetic properties measured using *m*(*H*) loops and FMR are summarized in **Table 1.**

The enhancement in $\alpha$ for increasing *c*-axis texture of $Bi_2Te_3$ can be attributed to a large spin-pumping effect modeled by the spin-mixing conductance, $g_{\uparrow\downarrow} = \frac{4\pi M_s t_{FM} \Delta\alpha}{\hbar \gamma}$, where $t_{FM}$ is the thickness of the FM layer and $\Delta\alpha$ is the enhancement in Gilbert damping and $\hbar$ is the reduced Plank's constant. The resulting $g_{\uparrow\downarrow}$ values increased with increasing crystalline *c*-axis orientation of $Bi_2Te_3$ and, were $2.10\times10^{-18}$, $2.38\times10^{-18}$ and $4.08\times10^{-18}$ m$^{-2}$ for the GBT/Py, DBT/Py and CBT/Py samples, respectively. The $g_{\uparrow\downarrow}$ values were calculated assuming the Gilbert damping enhancement in the $Bi_2Te_3$/Py samples are entirely due to spin-pumping. The loss of magnetization from interfacial diffusion of Ni from Py and spin-memory loss due to interfacial proximity-induced magnetization [13-19] may also play a role in the enhancement of $\alpha$ in $Bi_2Te_3$/Py samples. But these contributions towards enhancement in $\alpha$ could not be isolated because of complexity in these heterostructure material systems. However, the large enhancement in $\alpha$ with highly c-axis-oriented TIs is also observed in other TI/FM materials systems [27,33-37], including those which do not show interfacial diffusion (see Supporting Information Section S4). This suggests that for highly crystalline oriented TIs, the TI/FM heterostructures experience a giant enhancement in spin-pumping predominantly from the presence of robust TSS. The reduction in magnetization of the Py layer because of diffusion of Ni and reaction with $Bi_2Te_3$ was previously shown [25]. These results provide strong evidence of enhancement in SOC strength and topological properties in highly c-axis-oriented TI samples compared to disordered TIs.

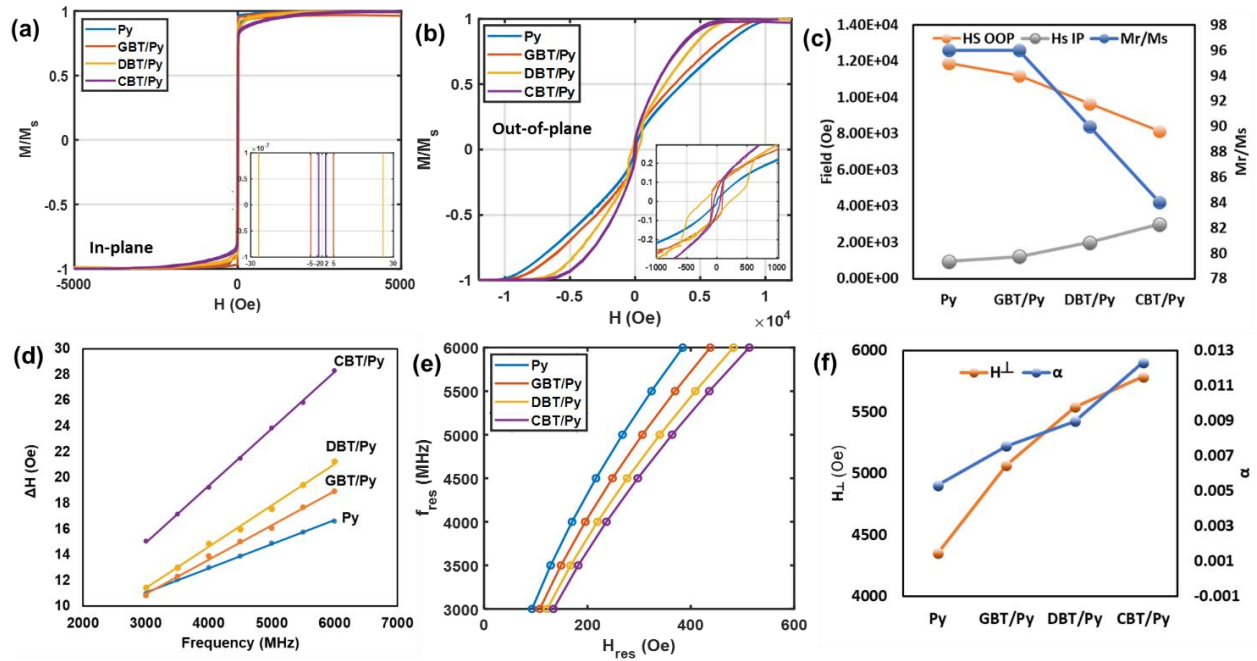

*Figure 3. Normalized m(H) loops measured: **a)** in-plane and **b)** out-of-plane for the GBT/Py, DBT/Py and CBT/Py samples. Inset: Expanded low-field regions showing enhanced $H_c$ for the $Bi_2Te_3$/Py samples compared to the Py control sample. **c)** Comparison of the saturation fields (IP and OP) and $M_r/M_s$ ratio clearly highlighting an increase in OP anisotropy with crystalline c-axis*

orientation of $Bi_2Te_3$. **d)** *FMR linewidth versus frequency for extraction of α.* **e)** *FMR resonance frequency versus field for extracting $H_\perp$ and $4\pi M_{eff}$.* **f)** *Visual comparison of α and $H_\perp$ extracted from d and e respectively. All measurements here were performed at 300 K.*

**Table 1.** Summary of room-temperature magnetic properties of the GBT/Py, DBT/Py and CBT/Py heterostructure samples.

| Sample | $H_s$ OOP (kOe) | $H_s$ IP (kOe) | $M_r/M_s$ (%) | $4\pi M_s$ (kOe) | $H_\perp$-FMR (kOe) | α |
|---|---|---|---|---|---|---|
| Py | 11.9 | 0.95 | 96 | 1.52 | 4.35 | 0.0053 |
| GBT/Py | 11.2 | 1.22 | 96 | 1.47 | 5.07 | 0.0075 |
| DBT/Py | 9.7 | 2.01 | 90 | 1.43 | 5.54 | 0.0089 |
| CBT/Py | 8.1 | 3.05 | 84 | 1.41 | 5.78 | 0.0123 |

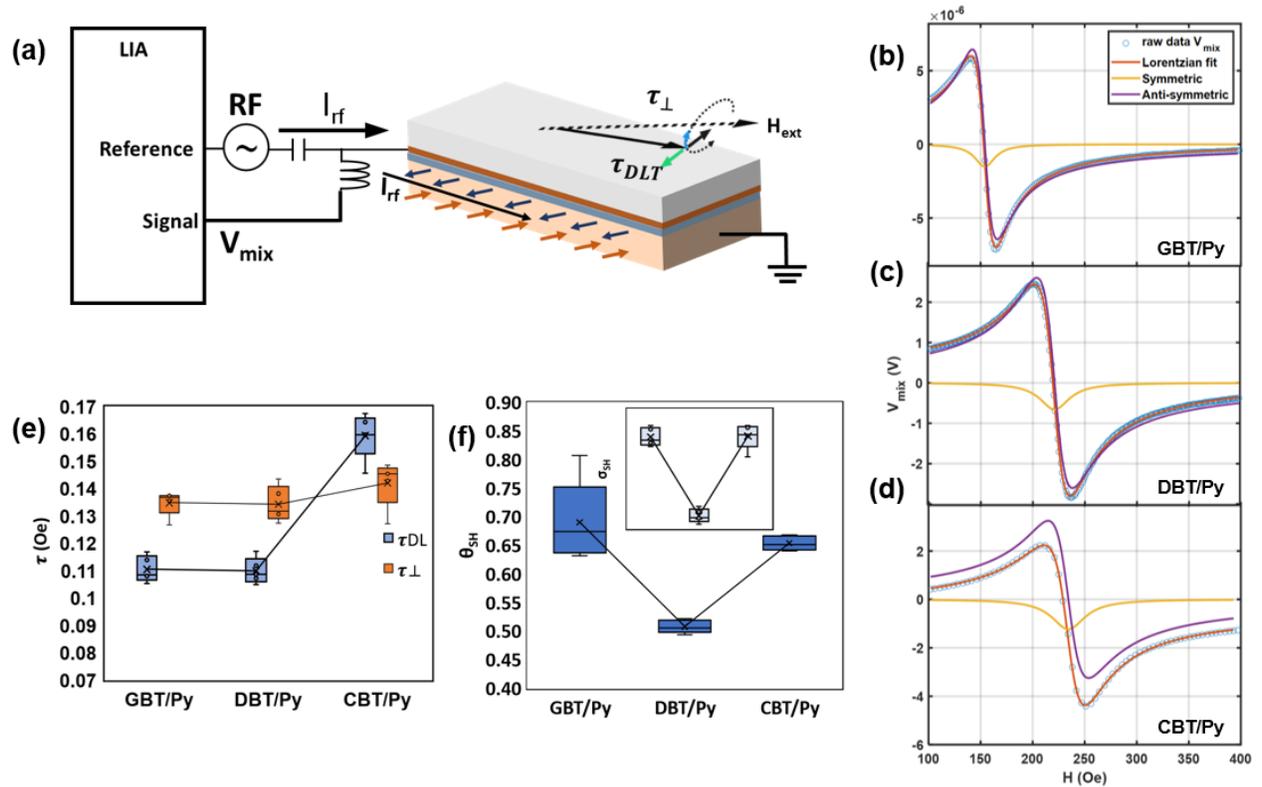

*Figure 4. a) Schematic for ST-FMR experimental setup with LIA and phase-locked RF current source. The brown and dark blue arrows signify up-spin and down-spin states, respectively. ST-FMR data and Lorentzian fitting for: b) GBT/Py, c) DBT/Py, and d) CBT/Py samples, measured at 4 GHz frequency. e) DL-SOT (blue) and Oersted plus FL-SOT for the GBT/Py, CBT/Py and DBT/Py samples extracted from b-d. f) $\theta_{SH}$ of the GBT/Py, CBT/Py and DBT/Py samples. Inset: $\sigma_{SH}$ of the GBT/Py, CBT/Py and DBT/Py samples.*

**2.4. Spin-Orbit Torque Properties of CBT/Py, DBT/Py and GBT/Py Samples.** The spin-orbit torque (SOT) characteristics were extracted from the symmetric and antisymmetric components of the fitted Lorentzian, as shown in Figures 4b-d, using the equations [43], $V_S = -\frac{I_{rf}\gamma \cos\theta_H}{4}\frac{dR}{d\theta_H}[\tau_{DL}\frac{1}{\Delta}F_{sym}]$ and $V_A = -\frac{I_{rf}\gamma \cos\theta_H}{4}\frac{dR}{d\theta_H}\tau_\perp \frac{\left(1+\frac{\mu_0 M_{eff}}{H_{ext}}\right)^{\frac{1}{2}}}{\Delta_{freq}}F_{asym}]$ (see Supporting Information Sections 3-5). Here, $I_{rf}$ is the RF current injected, $\theta_H$ is the in-plane angle of the external DC field relative to the injected RF current, $\frac{dR}{d\theta_H}$ is the derivative of the anisotropic magnetoresistance (AMR) relative to $\theta_H$, $\Delta_{freq}$ is the linewidth in the frequency domain, $\tau_{DL}$ is the damping-like DL-SOT, $\tau_\perp$, includes a combination the Oersted field torque and the field-like SOT (FL-SOT), $\gamma$ is the gyromagnetic ratio, $\mu_0$ is the permeability of vacuum and $\Delta H$ is the linewidth of the FMR signal. The $\tau_{DL}$ corresponds to the symmetric component of the Lorentzian, while the $\tau_\perp$ correspond to the antisymmetric components of the Lorentzian function [43]. The DL-SOT was the largest in the CBT/Py sample with a value of 0.16 Oe, compared to 0.11 Oe and 0.10 Oe in GBT/Py and DBT/Py, respectively, as shown in Figure 4e. The spin-Hall conductivity, $\sigma_{SH}$, which measures the spin current, $J_s$ generated from electric field, $E$ across the STFMR device is given by $\sigma_{SH} = \frac{J_s}{E} = \frac{\tau_{DL}M_s t_{FM}}{E}$. The average $\sigma_{SH}$ values shown in Figure 4f were calculated to be

$8.1 \times 10^4 \frac{\hbar}{2e}$, $5.5 \times 10^4 \frac{\hbar}{2e}$ and $8.1 \times 10^4 \frac{\hbar}{2e} \Omega^{-1} m^{-1}$, for GBT/Py, DBT/Py, and CBT/Py samples, respectively. Assuming negligible FL-SOT, the spin-charge current conversion efficiency measured by the spin-Hall angle given by, $\theta_{SH} = = (\frac{V_s}{V_a})(\frac{e\mu_0 M_s t_{FM} t_{TI}}{\hbar})\sqrt{1 + (\frac{4\pi M_{eff}}{H_{ext}})}$ (see Supporting Information Section S5). The values of $\theta_{SH}$ are calculated as 0.69, 0.51 and 0.65 for the GBT/Py, DBT/Py and CBT/Py samples, respectively. The $\sigma_{SH}$ and $\theta_{SH}$ are presented in Figure 4f which follow similar trends as expected. The GBT/Py and CBT/Py samples have a much larger $\sigma_{SH}$ and $\theta_{SH}$ compared to the disordered polycrystalline DBT/Py sample. This points towards a reduction in charge-spin current conversion efficiency with degradation in crystalline ordering possibly due to scattering of spin current in the randomly oriented crystalline TI domains. The GBT/Py sample, however, regains the spin-charge conversion efficiency possibly because of quantum confinement effect in the smaller grain size of the GBT sample [26, 27]. The symmetric Lorentzian in ST-FMR also includes a contribution from spin-pumping due the inverse spin-Hall effect (ISHE), which results in the $\alpha$ enhancement due to spin-pumping, as shown in the Figure 3d of the main text. However, because of the complexity of the interface in the $Bi_2Te_3$/Py samples, ISHE contribution to the symmetric component of FMR spectra could not be accurately isolated from the DL-SOT. Furthermore, the contribution of spin-pumping in Py-based heterostructures has been shown to be much smaller than the AMR component [44,45]. Hence, the ISHE components can be safely neglected from the calculations of DL-SOT and $\theta_{SH}$ for comparison of the samples.

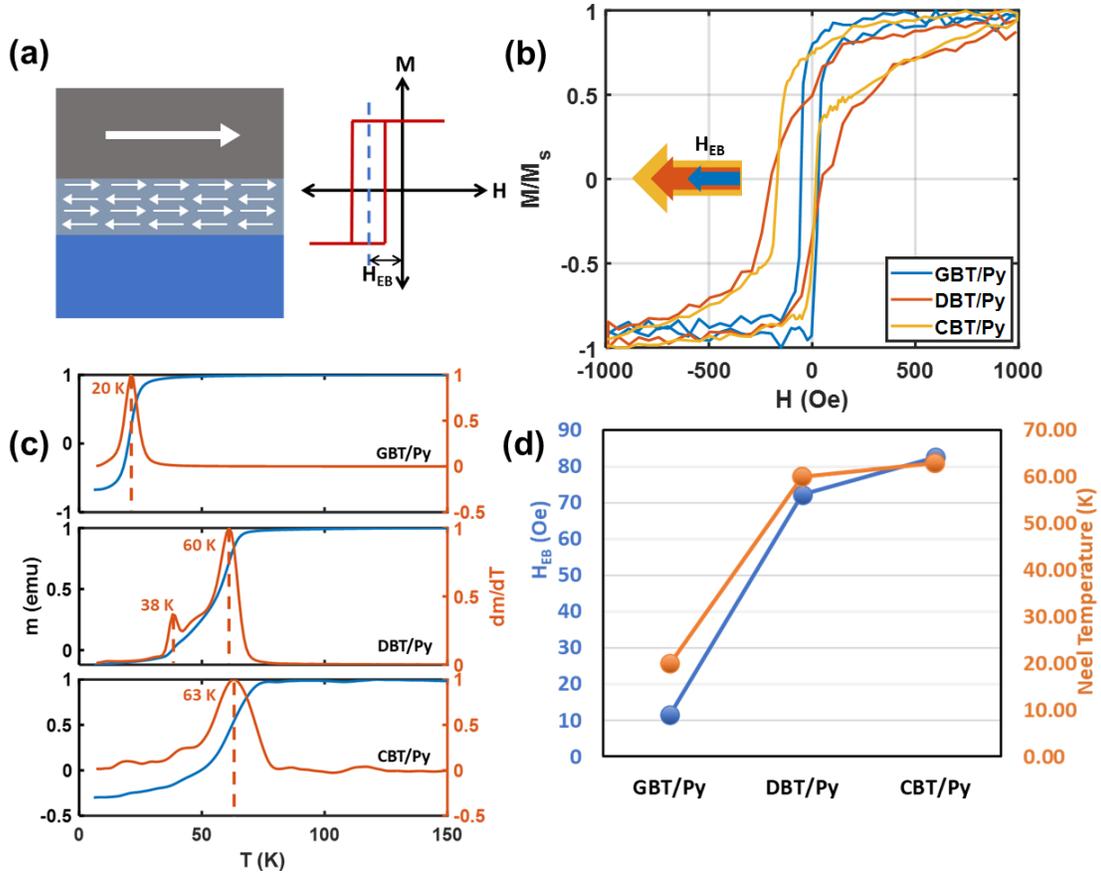

***Figure 5. a)*** *Schematic of the AFM-FM heterostructure materials system and exchange bias in m(H) loops.* ***b)*** *m(H) loops measured at 6 K under ZFC condition of the GBT/Py, DBT/Py and CBT/Py [25] samples showing spontaneous exchange bias. The size of the arrows qualitatively indicates the magnitude of shifts in the exchange bias.* ***c)*** *m(T) measurements of the GBT/Py, DBT/Py and CBT/Py samples and their derivatives for characterization of $T_N$.* ***d)*** *Exchange bias and $T_N$ values of the GBT/Py, DBT/Py and CBT/Py samples extracted from b and c. The m(H) and m(T) data presented for the CBT sample are similar to the ones in ref [25].*

**2.5. Effect of Crystalline Disorder in Interfacial Topological AFM Phase.** Creation of an interfacial AFM-ordered layer was reported in the interface of highly *c*-axis-oriented $Bi_2Te_3$/Py heterostructures [25]. That AFM ordering in the interfacial layer was found to exist because of the presence of the topological AFM compound, $NiBi_2Te_4$. Here, the effect of crystalline disorder of

Bi$_2$Te$_3$ on the AFM property of the Ni/Fe-diffused Bi$_2$Te$_3$ interface was also studied using zero-field-cooled (ZFC), $m(H)$ and $m(T)$ measurements at low temperatures as shown in Figure 5. Whereas the $m(H)$ measurements performed at 300 K are well-centered along the $H$-field axis, the $m(H)$ loops measured at 6 K shown in Figure 5b are significantly shifted off-center. This shift in the magnetic hysteresis loop is characteristic of spontaneous exchange bias that arises from an interfacial AFM-FM interaction given by $H_{EB} = \frac{J_{ex}}{4\pi M_s t_{FM}}$ [38-42], as illustrated in Figure 5a. Here, $J_{ex}$ is the interfacial AFM-FM exchange energy and $t_{FM}$ is the thickness of the ferromagnetic layer. As shown in Figures 4b,d, the CBT/Py with highly $c$-axis-oriented crystalline textured Bi$_2$Te$_3$ has the largest exchange bias of $H_{EB}$ = 83 Oe, while the DBT/Py sample with randomly oriented polycrystalline grains of Bi$_2$Te$_3$ has a slightly reduced exchange bias of $H_{EB}$ = 73 Oe. The exchange interaction strength in AFM materials is related to the Néel temperature, $T_N$ which were determined using ZFC $m(T)$ measurements [39,41] at a constant field of 50 Oe, as shown in Figure 5c. The CBT/Py and DBT/Py samples also show high values of $T_N$ = 63 and 60 K, respectively. It must also be noted that the large exchange bias and high $T_N$ in the disordered DBT/Py sample should also have significant contributions from the larger Ni and Fe interface concentration of 46% and 9% respectively, possibly causing the $H_{EB}$ and $T_N$ to be comparable to the highly $c$-axis-oriented CBT/Py sample. The DBT/Py sample also shows a secondary magnetic phase at 38 K also as observed from the smaller peak in the d$m$/d$T$ plot in Figure 5c which can also influence the $H_{EB}$ and $H_c$ in the sample at measurement temperature of 6 K. This possibly emerges due to the randomly oriented vdW crystalline domains affecting exchange interaction between the interfacial magnetic species. However, the highly disordered GBT/Py sample with granular Bi$_2$Te$_3$ had a large reduction in exchange bias to $H_{EB}$ = 12 Oe and $T_N$ = 20 K, clearly showing reduction in exchange interaction strength in the AFM interfacial layer. This follows from the much lower $\Delta m$ and hence lesser reaction between the diffused Ni and $Bi_2Te_3$ in the GBT/Py sample compared to the DBT/Py and CBT/Py samples. In addition to the spontaneous exchange bias, the $m(H)$ measurements also show a characteristic enhancement in coercive field ($H_c$) in all the $Bi_2Te_3$/Py samples, as shown in Figure 5b, due to frustrated magnetic moments at the interface [40]. These results indicate the persistence of exchange properties even in highly disordered TIs.

3. CONCLUSION

Interfacial and magnetic properties of sputtered-deposited TI/FM $Bi_2Te_3$/Py heterostructures were studied for varying crystalline structural disorder of the $Bi_2Te_3$. An interface layer was found to form because of diffusion of Ni and smaller amounts of Fe into $Bi_2Te_3$. The diffused Ni and Fe were found to undergo solid-state chemical reactions with $Bi_2Te_3$ promoted by the TSS electrons. With increasing crystalline *c*-axis-oriented texture of the $Bi_2Te_3$, the strengthening of topological property of $Bi_2Te_3$ led to an enhanced reaction between the diffused species and $Bi_2Te_3$, which was observed by a larger loss of Py magnetic moment. Increase in crystalline *c*-axis orientation of $Bi_2Te_3$ also resulted in a notable increase in OP magnetic anisotropy, Gilbert damping and spin-orbit torque as observed from $m(H)$ loop and FMR measurements. Interestingly, polycrystalline disordered $Bi_2Te_3$ sample had a reduced charge-spin current conversion efficiency possibly because of scattering of spins from polycrystalline grain boundaries. Whereas the samples with granular and highly c-axis-oriented $Bi_2Te_3$ had a comparable charge-spin current conversion efficiencies, which possibly resulted from quantum confinement effect in smaller crystalline grains and strong TSS, respectively. As such, this is expected to result in reduced spin-charge conversion efficiency. Furthermore, low temperature magnetization measurements showed surprising resilience of the topological property of $Bi_2Te_3$ as the AFM order persisted even in highly disordered granular $Bi_2Te_3$/Py heterostructures. However, for this sample the exchange interaction strength of the interfacial AFM phase was found to weaken significantly with the increase in structural disorder of $Bi_2Te_3$. This was verified by degradation in $H_{EB}$ and $T_N$ with increase in disorder of the $Bi_2Te_3$. These results open the path for further exploration of crystalline disorder in TIs and TI/FM interfaces. These thin film heterostructures can be easily grown using a CMOS-compatible sputtering process that can lead to realization of energy efficient spintronic devices.

## 4. EXPERIMENTAL SECTION

**4.1. Material Growth.** $Bi_2Te_3$ thin films of thickness 30 nm were grown by co-sputtering a composite $Bi_2Te_3$ target with Te target, using RF magnetron sputtering at 90 W and 20 W, respectively, with 4 mTorr Ar pressure on thermally oxidized Si substrates. The base pressure of the sputtering chamber was ~$8 \times 10^{-8}$ Torr. The GBT, DBT and CBT samples were grown with substrate maintained at 20°C, 160°C and 250°C, respectively. The 160°C DBT and 250°C CBT samples were further annealed at the growth temperatures inside the PVD process chamber in 45 mTorr pressure in Ar environment for 25 minutes. The CBT samples were gown using the same method as ref [25]. The samples were capped with 2 nm Al at room temperature before breaking vacuum which oxidizes to $AlO_x$ on exposure to atmosphere. For the magnetic and ST-FMR experiments, 20 nm Py and 3 nm $TiO_x$ capping were deposited at room temperature after deposition of $Bi_2Te_3$.

**4.2. XRD Characterization.** X-ray diffraction was collected using a background-free, highly collimated beam of Cu-Kα1 radiation (wavelength λ = 1.54056 Å). The X-rays were captured by a 2D charged-coupled device (CCD). The Bragg reflections were indexed according to the $Bi_2Te_3$ bulk hexagonal unit cell, as indicated by (*h*, *k*, -(*h*+*k*), *l*) where *h*, *k*, and *l* are the Miller indices [25].

**4.3. TEM and XEDS Characterization.** Samples for TEM investigations were prepared by focused ion beam milling (FIB) using a Ga+ ion source. Prior to TEM observation an additional cleaning procedure was performed by Ar-ion milling to reduce a surface amorphous layer and residual Ga from the FIB process. The TEM observations were performed using a Talos 200-FX (ThermoFiszher Scientific Inc.) TEM operated at an acceleration voltage of 200 kV [25]. EDS measurements were performed using a ChemiSTEM (ThermoFisher Scientific) and processing of the spectra was performed using Esprit 1.9 (Brucker Inc.) software [25].

**4.4. FMR Measurements.** FMR measurements of $\alpha$, $4\pi M_{eff}$, and $H_a$ were performed using a spin-torque FMR (ST-FMR) experimental setup. The analysis of the experiment is explained in Supporting Information Section S4. RF current is provided by a HP8350 RF source. A SR830 lock-in amplified (LIA) provides reference low-frequency AC for modulation phase-locked with the RF current. The LIA was used for detection of ST-FMR signal. The bias DC field is provided by an Fe-core electromagnet on a rotating stage with precise angular control. The reported ST-FMR experiment was performed at a 45° angle of the microstrip relative to the DC bias field. Control of the experiment and data acquisition was done using NI LabVIEW. The S11, S12 and impedance values were used to calculate the RF current, and the *E*-field using vector network analyzer (VNA). The total power lost in the electrical components, such as wires and connectors, were measured to be ~60%, hence, 40% of 8 dbm power was used for RF current calculations. FMR characteristics in the $Bi_2Te_3$/Py heterostructure samples were also extracted from the ST-FMR experiment spectra as shown in Figure S2. The devices were patterned using ion-milling process for the DBT/Py and CBT/Py samples and using lift-off process for the Py and GBT/Py samples. The FMR characteristics were extracted by fitting Lorentzian functions to the spectra [43], as shown in Figures S2a-d, using the equation, $V_{mix} = V_S F_{sym} + V_A F_{asym}$, which clearly shows broadening of FMR linewidth progressively from Py, GBT/Py, DBT/Py to CBT/Py samples. Here, $V_{mix}$ is the DC voltage output recorded in the LIA, $F_{sym} = \frac{\Delta H^2}{(\mu_0 H - \mu_0 H_{res})^2 + \Delta H^2}$ and $F_{asym} = \frac{\Delta H(\mu_0 H - \mu_0 H_{FMR})}{(\mu_0 H - \mu_0 H_{res})^2 + \Delta H^2}$ are the symmetric and antisymmetric components of the Lorentzian function, $\Delta H$ is the linewidth of the FMR signal, and $H_{res}$ is the FMR field. The ST-FMR measurements were performed at 45° angle relative to the external applied magnetic field. The values of $\Delta H$, $H_{FMR}$, $V_S$ and $V_A$ were extracted by fitting the ST-FMR signal using above equation for the analysis of $\alpha$, $4\pi M_s$, $H_\perp$ and $\theta_{SH}$ reported in the main text [25].

**4.5. Hysteresis Loop Measurements.** Magnetization $m(H)$ and $m(T)$ measurements were obtained using a Quantum Design MPMS XL-7 superconducting quantum interference device (SQUID) magnetometer [25]. Hysteresis loop $m(H)$ measurements were carried out at various temperatures between 6 and 300 K. The ZFC $m(T)$ measurements were obtained while increasing the temperature in an applied field of 50 Oe. Room temperature $m(H)$ measurements were taken using a vibrating sample magnetometer (VSM).

**Present Addresses**

†Dartmouth College, Thayer School of Engineering, Hanover, NH 03755

**Notes**

The authors declare no competing financial interests.

**Author Contributions**

The manuscript was written through contributions of all authors. All authors have given approval to the final version of the manuscript.

ACKNOWLEDGEMENT

We thank Charles Settens and MIT, Materials Research Laboratory for their help with XRD measurements. We thank Neville Sun and Mehdi Nasrollahpourmotlaghzanjani for help with VNA measurements. We also thank Ivan Lisenkov for his valuable input in understanding FMR experiments. Certain commercial equipments are identified in this paper to foster understanding, but such identification does not imply recommendation or endorsement by Northeastern University and AFRL.

**Funding Sources**

This work is partially supported by the U.S Army under grant no. W911NF20P0009, the NIH Award UF1NS107694 and by the NSF TANMS ERC Award 1160504. The work of DH and AF was partially supported by the National Science Foundation grant DMR-1905662 and the Air Force Office of Scientific Research award FA9550-20-1-0247. The work of KM was supported by Air Force Research Laboratory under AFRL/NEMO contract: FA8650-19-F-5403 TO3. Studies employing the Titan 60-300 TEM was performed at the Center for Electron Microscopy and Analysis (CEMAS) at The Ohio State University with support through Air Force contract FA8650-18-2-5295.

ABBREVIATIONS

AFM, Antiferromagnet; CBT, c-axis oriented Bi2Te3; DBT, Disordered Bi2Te3; DL-SOT, Dampin-like spin orbit torque; EDS, Energy-dispersive X-ray spectroscopy; FM, Ferromagnet; FMR, Ferrommagnetic resonance; GBT, Granular Bi2Te3; HRTEM, High resolution transmission electron microscopy; IP, In-plane; MBE, Molecular beam epitaxy; OP, Out-of-plane; QAH, Quantum anomalous hall; RF, Radio frequency; SOC, Spin orbit coupling; TI,

Topological insulator; TSS, Topological surface states; XRD, X-ray diffraction; XRR, X-ray reflectometry; ZFC, Zero field cooled.

REFERENCES

1. H Zhang, C.X. Liu, X.L. Qi, X. Dai, Z. Fang and S.C. Zhang, Topological insulators in $Bi_2Se_3$, $Bi_2Te_3$ and $Sb_2Te_3$ with a single Dirac cone on the surface. Nat. Phys. 5, 438–442 (2009).

2. Y. L. Chen, J. G. Analytis, J.-H. Chu, Z. K. Liu, S.-K. Mo, X. L. Qi, H. J. Zhang, D. H. Lu, X. Dai, Z. Fang, S. C. Zhang, I. R. Fisher, Z. Hussain, Z.-X. Shen, Experimental Realization of a Three-Dimensional Topological Insulator, $Bi_2Te_3$, Science, Vol 325, 5937 (2009).

3. Y. Zhang, K. He, C.Z. Chang, C.L. Song, L.L. Wang, X. Chen, J.F. Jia, Z. Fang, X. Dai, W.Y. Shan, S.Q. Shen, Q. Niu, X.L. Qi, S.C. Zhang, X.C. Ma and Q.K. Xue, Crossover of the three-dimensional topological insulator $Bi_2Se_3$ to the two-dimensional limit, Nat. Phys. 6, 584–588 (2010).

4. R. Yu, W. Zhang, H.J. Zhang, S.C. Zhang, X. Dai, Z. Fang, Quantized anomalous Hall effect in magnetic topological insulators. Science 329, 61–64 (2010).

5. C.Z. Chang, J. Zhang, X. Feng, J. Shen, Z. Zhang, M. Guo, K. Li, Y. Ou, P. Wei, L.-L. Wang, Z.-Q. Ji, Y. Feng, S. Ji, X. Chen, J. Jia, X. Dai, Z. Fang, S.-C. Zhang, K. He, Y. Wang, L. Lu, X.-C. Ma, Q.-K. Xue, Experimental observation of the quantum anomalous Hall effect in a magnetic topological insulator. Science 340, 167–170 (2013).

6. C.Z. Chang, W. Zhao, D. Y. Kim, H. Zhang, B. A. Assaf, D. Heiman, S.-C. Zhang, C. Liu, M. H. Chan, J. S. Moodera, High-precision realization of robust quantum anomalous Hall state in a hard ferromagnetic topological insulator. Nat. Mater. 14, 473–477 (2015).

7.  Y. Tokura, K. Yasuda and A. Tsukazaki, Magnetic Topological Insulators, Nat. Rev. Phys. 1, 126–143 (2019).

8.  C. Liu, Y. Wang, H. Li, Y. Wu, Y. Li, J. Li, K. He, Y. Xu, J. Zhang, Y. Wang, Robust axion insulator and Chern insulator phases in a two-dimensional antiferromagnetic topological insulator, Nat. Mater. 19, 522–527 (2020).

9.  W. Wang, Y. Ou, C. Liu, Y. Wang, K. He, Q.K. Xue, W. Wu, Direct evidence of ferromagnetism in a quantum anomalous Hall system, Nat. Phys. 14, 791–795 (2018).

10. J. Teng, N. Liu, and Y. Li, Mn-doped topological insulators: a review, J. Semicond. 40, 081507 (2019).

11. A. Tcakaev, V. B. Zabolotnyy, C. I. Fornari, P. Rüßmann, T. R. F. Peixoto, F. Stier, M. Dettbarn, P. Kagerer, E. Weschke, E. Schierle, P. Bencok, P. H. O. Rappl, E. Abramof, H. Bentmann, E. Goering, F. Reinert, and V. Hinkov, Incipient antiferromagnetism in the Eu-doped topological insulator $Bi_2Te_3$, Phys. Rev. B 102, 184401 (2020).

12. Y. Ni, Z. Zhang, I. C. Nlebedim, R. L. Hadimani, G. Tuttle, D. C. Jiles, Ferromagnetism of magnetically doped topological insulators in $Cr_xBi_{2-x}Te_3$ thin films, J Appl. Phys. 117, 17C748 (2015).

13. F. Katmis, V. Lauter, F.S. Nogueira, B.A. Assaf, M.E. Jamer, P. Wei, B. Satpati, J.W. Freeland, I. Eremin, D. Heiman, P. Jarillo-Herrero, J.S. Moodera, A high-temperature ferromagnetic topological insulating phase by proximity coupling, Nat. 533, 513–516 (2016).

14. X. Che, K. Murata, L. Pan, Q.L He, G. Yu, Q. Shao, G. Yin, P. Deng, Y. Fan, B. Ma, X. Liang, B. Zhang, X. Han, L. Bi, Q.H. Yang, H. Zhang, K. L. Wang, Proximity-Induced Magnetic

Order in a Transferred Topological Insulator Thin Film on a Magnetic Insulator, ACS Nano 12, 5042−5050 (2018).

15. C. Lee, F. Katmis, P. Jarillo-Herrero, J.S. Moodera and N. Gedik, Direct measurement of proximity-induced magnetism at the interface between a topological insulator and a ferromagnet, Nat. Comm. 7, 12014 (2016).

16. W.Y. Choi, J. H. Jeon, H.W. Bang, W. Yoo, S.K. Jerng, S.H. Chun, S. Lee, M.H. Jung, Proximity-Induced Magnetism Enhancement Emerged in Chiral Magnet MnSi/Topological Insulator $Bi_2Se_3$ Bilayer, Adv. Quant. Tech. 4, 2000124 (2021).

17. J. A. Hutasoit, T.D. Stanescu, Induced spin texture in semiconductor/topological insulator heterostructures, Phys. Rev. B 84, 085103 (2011).

18. J.M. Marmolejo-Tejada, K. Dolui, P. Lazić, P.H. Chang, S. Smidstrup, D. Stradi, K. Stokbro, and B. K. Nikolic, Proximity Band Structure and Spin Textures on Both Sides of Topological Insulator/Ferromagnetic-Metal Interface and Their Charge Transport Probes, Nano Lett. 17, 5626−5633 (2017).

19. I. Zutic, A. Matos-Abiague, B. Scharf, H. Dery, K. Belashchenko, Proximitized Materials, Mater. Today 22, 85 (2019).

20. J. Li, Y. Li, S. Du, Z. Wang, B.L. Gu, S.C. Zhang, K. He, W. Duan, Y. Xu, Intrinsic magnetic topological insulators in van der Waals layered $MnBi_2Te_4$-family materials, Sci. Adv. 5, eaaw5685 (2019).

21. Z. Li, J. Li, K. He, X. Wan, W. Duan, Y. Xu, Tunable interlayer magnetism and band topology in van derWaals heterostructures of MnBi$_2$Te$_4$-family materials, Phys. Rev. B 102, 081107(R) (2020).

22. L. A. Walsh, C. M. Smyth, A. T. Barton, Q. Wang, Z. Che, R. Yue, J. Kim, M. J. Kim, R. M. Wallace, and C. L. Hinkle, Interface Chemistry of Contact Metals and Ferromagnets on the Topological Insulator Bi2Se3, J. Phys. Chem. C, 121, 23551-23563 (2017).

23. K. Ferfolja, M. Fanetti, S. Gardonio, M. Panighel, I. Pis, S. Nappini and M. Valant, A cryogenic solid-state reaction at the interface between Ti and the Bi$_2$Se$_3$ topological insulator, J. Mater. Chem. C 8, 11492-11498 (2020).

24. G. Li, C. Felser, Heterogeneous catalysis at the surface of topological materials, Appl. Phys. Lett. 116, 070501 (2020).

25. N. Bhattacharjee, K. Mahalingam, A. Fedorko, V. Lauter, M. Matzelle, B. Singh, A. Grutter, A. Will-Cole, M. Page, M. McConney, R. Markiewicz, A. Bansil, D. Heiman, and N.X. Sun, Topological Antiferromagnetic Van der Waals Phase in Topological Insulator/Ferromagnet Heterostructures Synthesized by a CMOS-Compatible Sputtering Technique, Adv. Mater., 2108790 (2022).

26. M. DC, R. Grassi, J. Y. Chen, M. Jamali, D. R. Hickey, D. Zhang, Z. Zhao, H. Li, P. Quarterman, Y. Lv, M. Li, A. Manchon, K. A. Mkhoyan, T. Low & J. P. Wang, Room-temperature high spin-orbit torque due to quantum confinement in sputtered Bi$_x$Se$_{(1-x)}$ films, Nature Mater. volume 17, 800 (2018).

27. M DC, T. Liu, J. Y. Chen, T. Peterson, P. Sahu, H. Li, Z. Zhao, M. Wu, and J. P. Wang, Room-temperature spin-to-charge conversion in sputtered bismuth selenide thin films via spin pumping from yttrium iron garnet, Appl. Phys. Lett. 114, 102401 (2019).

28. Q. Guo, Yu Wu, L. Xu, Y. Gong, Y. Ou, Y. Liu, L. Li, Y. Yan, G. Han, D. Wang, L. Wang, S. Long, B. Zhang, X. Cao, S. Yang, X. Wang, Y. Huang, T. Liu, G. Yu, K. He and J. Teng, Electrically Tunable Wafer-Sized Three-Dimensional Topological Insulator Thin Films Grown by Magnetron Sputtering, Chin. Phys. Lett. 37, 057301 (2020).

29. Qi.X. Guo, Z.X. Ren, Y.Y. Huang, Z.C. Zheng, X.M. Wang, W. He, Z.D. Zhu, and J. Teng, Effects of post-annealing on crystalline and transport properties of $Bi_2Te_3$ thin films, Chin. Phys. B, Vol. 30(6): 067307 (2021).

30. P. Corbae, S. Ciocys, D. Varjas, E. Kennedy, S. Zeltmann, M. Molina-Ruiz, S. Griffin, C. Jozwiak, Z. Chen, L.W. Wang, A. M. Minor, M. Scott, A. G. Grushin, A. Lanzara, F. Hellman, Evidence for topological surface states in amorphous $Bi_2Se_3$, arXiv:1910.13412 [cond-mat.mtrl-sci] (2021).

31. D. Yia, J. Liub, S.L. Hsua, L. Zhang, Y. Choig, J.W. Kimg, Z. Chena, J. D. Clarksona, C. R. Serraoa, E. Arenholzh, P. J. Ryang, H. Xuf, R. J. Birgeneaua and R. Ramesh, Atomic-scale control of magnetic anisotropy via novel spin–orbit coupling effect in $La_{2/3}Sr_{1/3}MnO_3/SrIrO_3$ superlattices, PNAS, Vol. 113, No. 23 (2016),

32. Y. G. Semenov, X. Duan and K. W. Kim, Electrically controlled magnetization in ferromagnet-topological insulator heterostructures, Phys. Rev. B 86, 161406(R) (2012).

33. T. Liu, J. Kally, T. Pillsbury, C. Liu, H. Chang, J. Ding, Y. Cheng, M. Hilse, R. Engel-Herbert, A. Richardella, N. Samarth and M. Wu, Changes of Magnetism in a Magnetic Insulator due to Proximity to a Topological Insulator, Phys Rev Lett. 125, 017204 (2020).

34. M. Jamali, J. S. Lee, J. S. Jeong, F. Mahfouzi, Y. Lv, Z. Zhao, B. K. Nikolić, K. A. Mkhoyan, N. Samarth, and J.P. Wang, Giant Spin Pumping and Inverse Spin Hall Effect in the Presence of Surface and Bulk Spin−Orbit Coupling of Topological Insulator $Bi_2Se_3$, Nano Lett., 15, 10, 7126 (2015).

35. H. Wang, J. Kally, C. Şahin, T. Liu, W. Yanez, E. J. Kamp, A. Richardella, M. Wu, M. E. Flatté, and N. Samarth, Fermi level dependent spin pumping from a magnetic insulator into a topological insulator, Phys. Rev. Res. 1, 012014(R) (2019).

36. Y.S. Hou and R.Q. Wu, Strongly Enhanced Gilbert Damping in 3d Transition-Metal Ferromagnet Monolayers in Contact with the Topological Insulator $Bi_2Se_3$, Phys. Rev. Appl. 11, 054032 (2019).

37. T. Chiba, A. O. Leon, and T. Komine, Voltage-control of damping constant in magnetic-insulator/topological-insulator bilayers, Appl. Phys. Lett. 118, 252402 (2021).

38. M. Li, C. Z. Chang, B. J. Kirby, M. E. Jamer, W. Cui, L. Wu, P. Wei, Y. Zhu, D. Heiman, J. Li, and J. S. Moodera, Proximity-Driven Enhanced Magnetic Order at Ferromagnetic-Insulator–Magnetic-Topological-Insulator Interface, Phys. Rev. Lett. 115, 087201 (2015).

39. J. K. Murthy, P. S. Anil Kumar, Interface-induced spontaneous positive and conventional negative exchange bias effects in bilayer $La_{0.7}Sr_{0.3}MnO_3/Eu_{0.45}Sr_{0.55}MnO_3$ heterostructures, Sci. Rep. 7, 6919 (2017)

40. C. Leighton, J. Nogués, B. J. Jönsson-Åkerman, I. K. Schuller, Coercivity Enhancement in Exchange Biased Systems Driven by Interfacial Magnetic Frustration, Phys. Rev. Lett. 84, 3466 (2000).

41. T. Maity, S. Goswami, D. Bhattacharya, S. Roy, Superspin Glass Mediated Giant Spontaneous Exchange Bias in a Nanocomposite of $BiFeO_3-Bi_2Fe_4O_9$, Phys. Rev. Lett. 110, 107201 (2013).

42. J. Liu, A. Singh, Y. Yang, F. Liu, A. Ionescu, B. Kuerbanjiang, C. H. W. Barnes, T. Hesjedal, Exchange Bias in Magnetic Topological Insulator Superlattices Nano Lett. 20, 5315 (2020).

43. Y. Wang, R. Ramaswamy and H. Yang, FMR-Related Phenomena in Spintronic Devices, J. Phys. D: Appl. Phys. 51, 273002 (2018).

44. K. Kondou, H. Sukegawa, S. Kasai, S. Mitani, Y. Niimi and Y. C. Otani, Influence of Inverse Spin Hall Effect in Spin-Torque Ferromagnetic Resonance Measurements, Appl. Phys. Exp. 9, 023002 (2016).

45. F. Bonell, M. Goto, G. Sauthier, J. F. Sierra, A. I. Figueroa, M. V. Costache, S. Miwa, Y. Suzuki, and S. O. Valenzuela, Control of Spin–Orbit Torques by Interface Engineering in Topological Insulator Heterostructures, Nano Lett., 20, 8, 5893 (2020).

**Supplementary Materials: Supporting Information for Effects of Crystalline Disorder on Interfacial and Magnetic Properties of Sputtered Topological Insulator/Ferromagnet Heterostructures**

**S1. Grain size calculations of Bi$_2$Te$_3$ Samples**

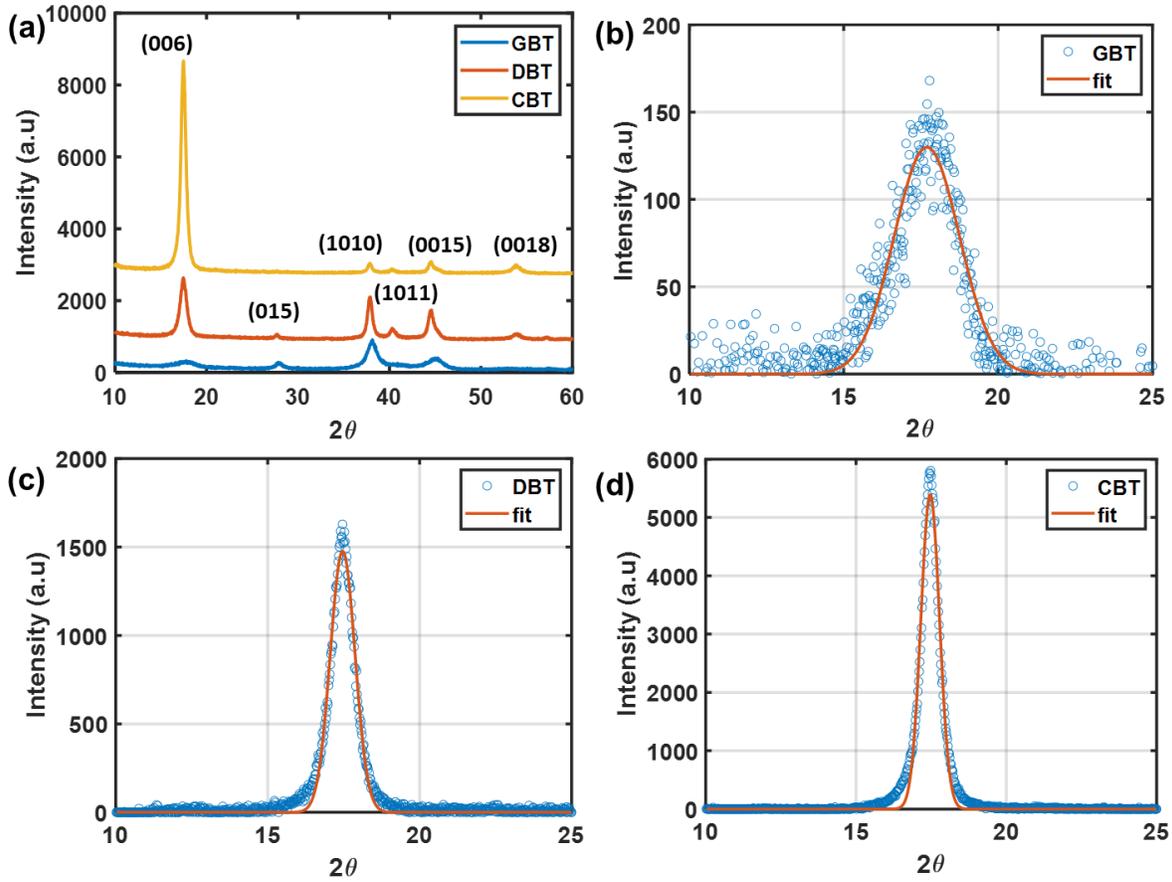

*Figure B1. a) GIXD plots for GBT, DBT and CBT samples. Representative Gaussian fitting of the GIXD data for b) GBT, c) DBT and d) CBT samples, used for extracting the FWHM.]*

As shown in Fig. S1a, grazing-angle XRD (GIXD) measured at a small incidence angle (~1º) clearly show the presence of comparable diffraction peak intensities from ($hkl$) = ($00l$), ($10l$) and ($11l$) orientations in both the GBT and DBT samples. The comparable intensities suggest the samples are disordered polycrystalline in nature, where $h$, $k$ and $l$ are the miller indices. The CBT sample had clear (*00l*) orientation that was determined from both symmetric-XRD (main text Fig. 1e) and GIXD measurements, showing that the CBT thin film is highly textured and oriented along the crystalline *c*-axis. Crystallite grain sizes for the sputter-grown $Bi_2Te_3$ samples were determined using the Scherrer equation, $d_{grain} = \frac{K\lambda}{FWHM \cos\theta}$. Here, $d_{grain}$ is the grain size, $K$ is the shape factor assumed to be 0.9, $\lambda$=1.54 Å is the wavelength of the X-ray source, $\theta$ is the Bragg angle and FWHM is the full-width at half-maximum of the peaks. The FWHM parameters of the peaks are extracted by fitting a Gaussian function to the raw data for the GIXD peaks as shown in Fig. S1. The average calculated grain sizes for the samples are 8.2 nm, 16.4 nm and 18.7 nm for GBT, DBT and CBT, respectively.

**S2. Summary of Cross-Sectional Atomic% in GBT/Py, DBT/Py and CBT/Py Samples**

**Table S1. Average atomic % of elements along the cross section of the GBT/Py, DBT/Py and CBT/Py samples, measured using cross-sectional EDS.**

| Sample | Element | $Bi_2Te_3$ | $Ni$-$Bi_2Te_3$ | Py* | Py |
|---|---|---|---|---|---|
| GBT/Py | Bi | 39 | 20 | 2 | 0 |
|  | Te | 57 | 39 | 6 | 0 |
|  | Ni | 4 | 38 | 68 | 81 |
|  | Fe | 0 | 3 | 25 | 19 |
| DBT/Py | Bi | 39 | 18 | 2 | 0 |
|  | Te | 57 | 26 | 4 | 0 |
|  | Ni | 3 | 47 | 74 | 81 |
|  | Fe | 0 | 9 | 19 | 19 |
| CBT/Py | Bi | 39 | 22 | 3 | 0 |
|  | Te | 60 | 34 | 5 | 0 |
|  | Ni | 0 | 39 | 70 | 81 |
|  | Fe | 0 | 4 | 22 | 19 |

## S3. FMR Spectra for Bi$_2$Te$_3$/Py Samples

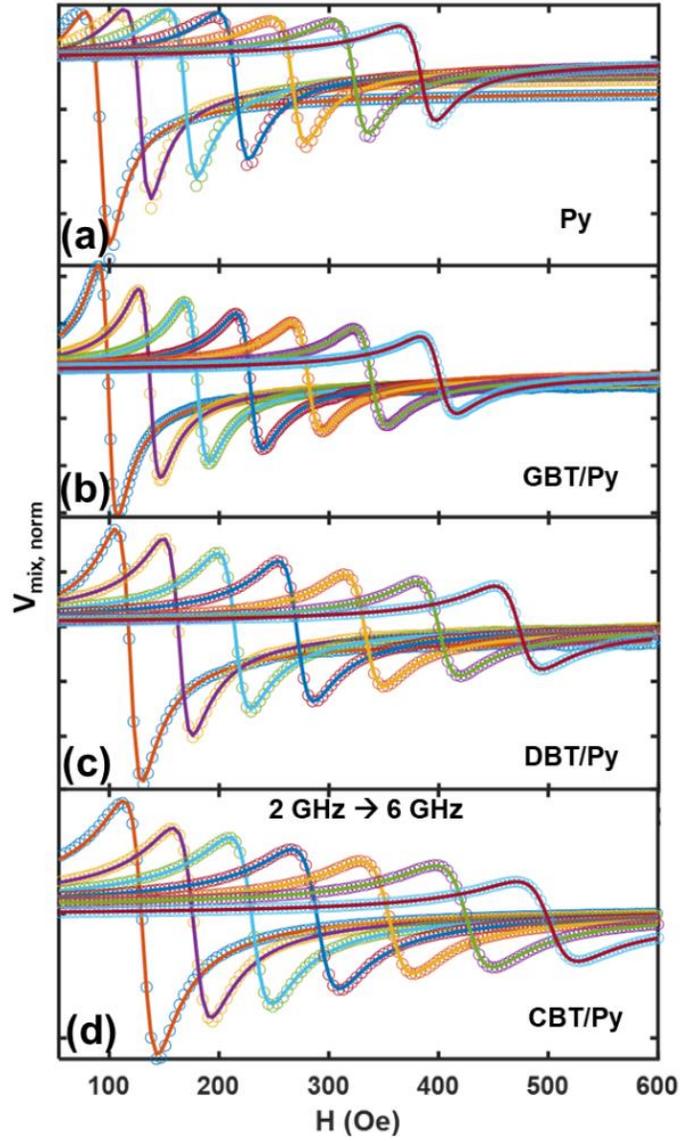

*Figure B2. Representative FMR spectra for the a) Py, b) GBT/Py, c) DBT/Py and d) CBT/Py samples, measured using ST-FMR experimental setup for extracting Gilbert damping and interfacial magnetic anisotropy field.*

FMR characteristics in the Bi$_2$Te$_3$/Py heterostructure samples were extracted from fitting Lorentzian functions to the ST-FMR experiment spectra as shown in Fig. S2. The devices were patterned using ion-milling process for the DBT/Py and CBT/Py samples and using lift-off process for the Py and GBT/Py samples (see Methods for details).

## S4. Calculation of Anisotropic Magnetoresistance in Bi2Te3/Py Samples

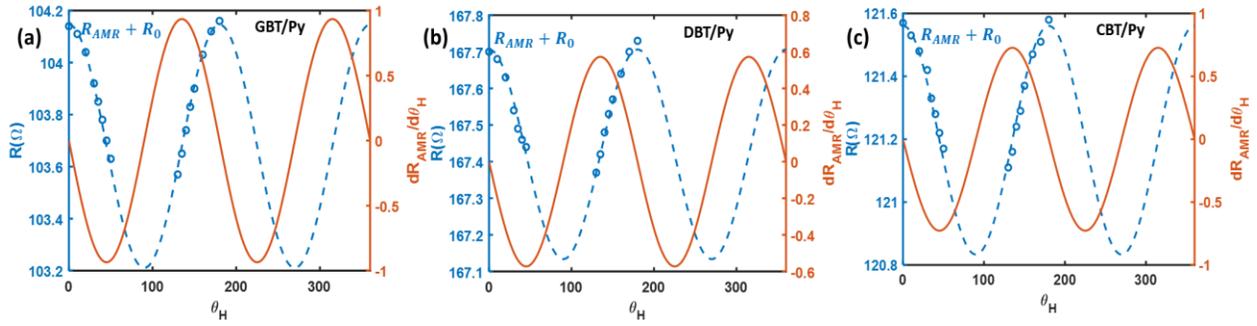

*Figure B3. Angular dependence of resistance of a) GBT/Py, b) DBT/Py and c) CBT/Py relative to the in-plane external field of 200 Oe, which was used to calculate AMR. The blue dots represent raw data, and the blue dashed curve is the theoretical fit.*

Anisotropic magnetoresistance (AMR), which is the angular dependence of the resistivity of the material in an applied DC magnetic field, is the primary component of the ST-FMR signal. The AMR of a rectangular microstrip can be characterized by in-plane angular-dependent measurement of resistance and is given by, $R = R_{AMR} \cos^2 \theta_H + R_0$. Here, $R$ is the total resistance of the microstrip, $R_{AMR}$ is the AMR component of the resistance and $R_0$ is the resistance at 90° angle. Figure S2 shows the resistance measured for the $Bi_2Te_3$/Py samples GBT, DBT and CBT at various angles in an in-plane magnetic field of 200 Oe. From fitting resistance measurements with the above equation, we obtained the average AMR values for the samples as 0.93 Ω, 0.57 Ω and 0.72 Ω, respectively. Further, taking a derivative of the above relation, we obtain the $dR/d\theta_H$ plots that are used in Eq. (3) in the main text for calculation of SOT field values using ST-FMR.

## S5. Spin-Hall Angle Analysis of Bi$_2$Te$_3$/Py Samples

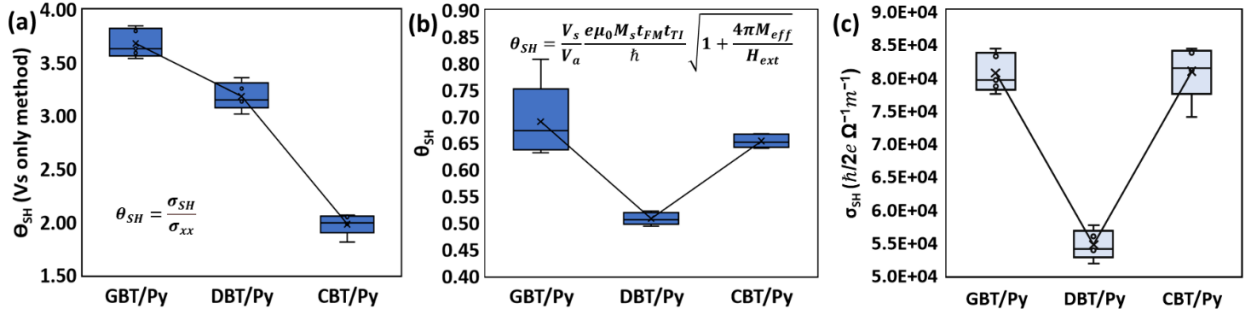

*Figure S4. Trend of a) θ$_{SH}$ calculated using the Vs-only method, b) θ$_{SH}$ calculated using the V$_s$/V$_a$ method and c) σ$_{SH}$ for the GBT/Py, DBT/Py and CBT/Py samples.*

Spin-Hall angle ($\theta_{SH}$) is the standard measure of spin-charge conversion efficiency in FM/normal metal (NM) heterostructure systems (Wang et al., J. Phys. D: Appl. Phys. **51**, 273002, 2018). In most cases, the $\theta_{SH}$ can be accurately determined from ST-FMR experiments using the Vs-only method, $\theta_{SH} = \frac{\sigma_{SH}}{\sigma_{xx}}$, where $\sigma_{xx}$ is the longitudinal conductivity of the NM thin film. The average $\sigma_{SH}$ values shown in Figure S4c are $8.1 \times 10^4 \frac{\hbar}{2e}$, $5.5 \times 10^4 \frac{\hbar}{2e}$ and $8.1 \times 10^4 \frac{\hbar}{2e} \Omega^{-1}m^{-1}$, for GBT/Py, DBT/Py, and CBT/Py samples, respectively. The determination of $\sigma_{xx}$ for the GBT, DBT and CBT thin films (with AlO$_x$ cap) without Py deposition was found to be $2.80 \times 10^{-4} \Omega^{-1}m^{-1}$, $1.72 \times 10^{-4} \Omega^{-1}m^{-1}$, and $4.08 \times 10^{-4} \Omega^{-1}m^{-1}$, respectively. Using the above relation, average $\theta_{SH}$ values obtained were 3.68, 3.18 and 1.98 for the GBT/Py, DBT/Py and CBT/Py, respectively, as shown in Fig. S4a. However, the Vs-only method assumes no interfacial diffusion, which leads to a non-physical result in these TI/FM materials systems. The resistance in the TI layer is certainly altered by the diffusion of Ni and Fe from Py and reaction with Bi$_2$Te$_3$. Moreover, the trend of $\theta_{SH}$ obtained from the Vs-only method does not follow the $\sigma_{SH}$ trend, which also means the $\theta_{SH}$ values obtained using the Vs-only method is incorrect. For a better estimation of the $\theta_{SH}$ parameter in these TI/FM systems with a complex interfacial morphology, the Vs/Va method given by, $\theta_{SH} = (\frac{V_s}{V_a})(\frac{e\mu_0 M_s t_{FM} t_{TI}}{\hbar})\sqrt{1 + (\frac{4\pi M_{eff}}{H_{ext}})}$, should be a more reliable method. This method assumes the antisymmetric component is completely from the Oersted field. Given the comparable values of $\tau_\perp$ for the GBT/Py, DBT/Py and CBT/Py (see Main Text Figure 4e), the FL-SOT can be safely assumed to be negligibly small compared to Oersted torque. Using the Vs/Va method, the average $\theta_{SH}$ for the GBT/Py, DBT/Py and CBT/Py were 0.69, 0.51 and 0.65, respectively, as shown in Fig. S4b. As expected, values also follow the trend of $\sigma_{SH}$ for the samples. This clearly suggests reduction in spin-charge conversion efficiency for the polycrystalline disordered TI sample DBT/Py.

## S6. Crystalline Disorder-Dependent Spin-Orbit Coupling Effects in TI/FM Heterostructures

### S6.1. Crystalline Disorder-Dependent Magnetic Properties of $Bi_2Te_3$/CoFeB

The crystalline disorder-dependent, room-temperature magnetic properties of TI/FM were also studied for $Bi_2Te_3$/CoFeB (CFB) heterostructure samples, as shown in Fig. S4. The FMR spectra show a giant enhancement in $\alpha$ and $H_\perp$ for the CBT/CFB samples with high $c$-axis orientation. The measurements were performed using a broad-band FMR setup for comparison with the $Bi_2Te_3$/Py samples. The extracted $\alpha$ values from FMR measurements in Fig. S4b were found to be 0.015, 0.026 and 0.283 for CFB, GBT/CFB and CBT/CFB, respectively. These increasing values for increasing order confirm the giant enhancement in SOC and strong TSS effects in highly $c$-axis oriented TI compared to highly disordered TI materials. Further, Kittel equation fits to the resonance frequency versus field in Fig. S4c give estimates for $H_\perp$ of -13.2 Oe, 100.7 Oe and 1485 Oe, for the respective sample revealing a giant enhancement in PMA field for the highly $c$-axis oriented $Bi_2Te_3$ sample. This clearly shows a large enhancement in interfacial PMA field in highly

*c*-axis oriented CBT/CFB samples compared to a higher in-plane anisotropy of the disordered GBT/CFB samples. The FMR measurements are also supported by the increase in $M_r/M_s$ ratio in the $m(H)$ measurements, shown in Fig. S4d. The $M_r/M_s$ values are 86%, 70% and 66% for the CFB, GBT/CFB and CBT/CFB samples, respectively, which shows an enhanced OP magnetic anisotropy for the highly oriented CBT/CFB sample compared to the disordered GBT/CFB sample. Table S2 summarizes the magnetic properties of the $Bi_2Te_3$/CoFeB samples obtained from the FMR and *m(H)* loop measurements.

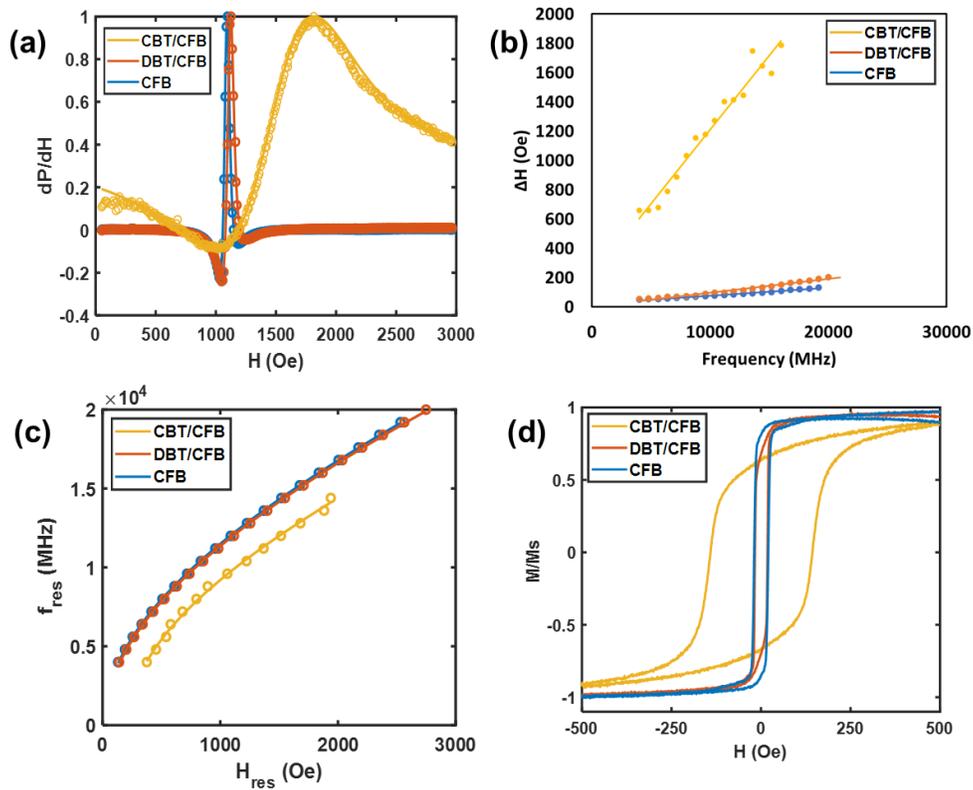

*Figure B5. a) Representative normalized FMR spectra for CFB, GBT/CFB and CBT/CFB at 6 GHz showing broadening of linewidth. b) FMR linewidth versus frequency plots and linear fit for extracting Gilbert damping in GBT/Py, CBT/Py and DBT/Py samples. c) FMR frequency versus*

*field used for extracting $4\pi M_s$ and $H_\perp$ in GBT/Py, CBT/Py and DBT/Py samples. d) m(H) measurement for the GBT/Py, CBT/Py and DBT/Py samples.*

**Table S2. Magnetic Properties of Bi$_2$Te$_3$/CoFeB Samples**

| Sample | $\alpha$ | $4\pi M_s$ (kOe) | $H_\perp$ (Oe) | $M_r/M_s$ (%) |
|---|---|---|---|---|
| CFB | 0.01484 | 16.01 | -13.17 | 86 |
| GBT/CFB | 0.02604 | 15.89 | 100.74 | 70 |
| CBT/CFB | 0.28336 | 14.51 | 1485.39 | 66 |

### S6.2. Crystalline Disorder-Dependent Magnetic Properties of YIG/Bi$_2$Te$_3$

The effect of TSS for giant enhancement in $\alpha$ was verified using Y$_3$Fe$_5$O$_{12}$ (YIG) as the FM layer. Disordered GBT and highly *c*-axis oriented CBT thin films of thickness 30 nm were grown on GGG (111)/YIG (444) (70 nm) substrates using the same deposition conditions mentioned in the main text. YIG is a chemically stable rare-earth garnet material, and because it is an oxide compound it is not expected to experience diffusion across the interface with TI (Chen et al., Appl. Phys. Lett. **114**, 031601, 2019). This was verified by *m(H)* measurements of YIG and YIG/GBT samples that did not show any reduction in magnetic moment, as shown in Fig. S4c, unlike with Py. FMR measurements for the YIG/Bi$_2$Te$_3$ samples revealed a similar trend as the Bi$_2$Te$_3$/FM samples and showed a giant enhancement in FMR linewidth and $\alpha$ for YIG/GBT heterostructures, as shown in Figs. S5b,d. The $\alpha$ values extracted from the linear fitting of $\Delta H$ versus resonance frequency plots in Figure S3d are $8.43 \times 10^{-4}$, $2.83 \times 10^{-3}$ and $7.09 \times 10^{-3}$ for increasing crystallinity. This clearly shows a giant enhancement in Gilbert damping in the YIG/CBT heterostructure sample as a result of the TSS in the highly *c*-axis oriented Bi$_2$Te$_3$. Fitting the Kittel equation also revealed changes in $\frac{\gamma}{2\pi}$ values from 2.81 for YIG to 2.83 and 2.84 for YIG/GBT and YIG/CBT samples, respectively. These changes in $\frac{\gamma}{2\pi}$ values provide strong evidence of interaction between the TSS and magnetic moments in YIG that are enhanced in the YIG/CBT sample (Liu et al., Phys. Rev. Lett. **125**, 017204, 2020). Table S3 summarizes the magnetic properties of the YIG/$Bi_2Te_3$ samples obtained from FMR measurements.

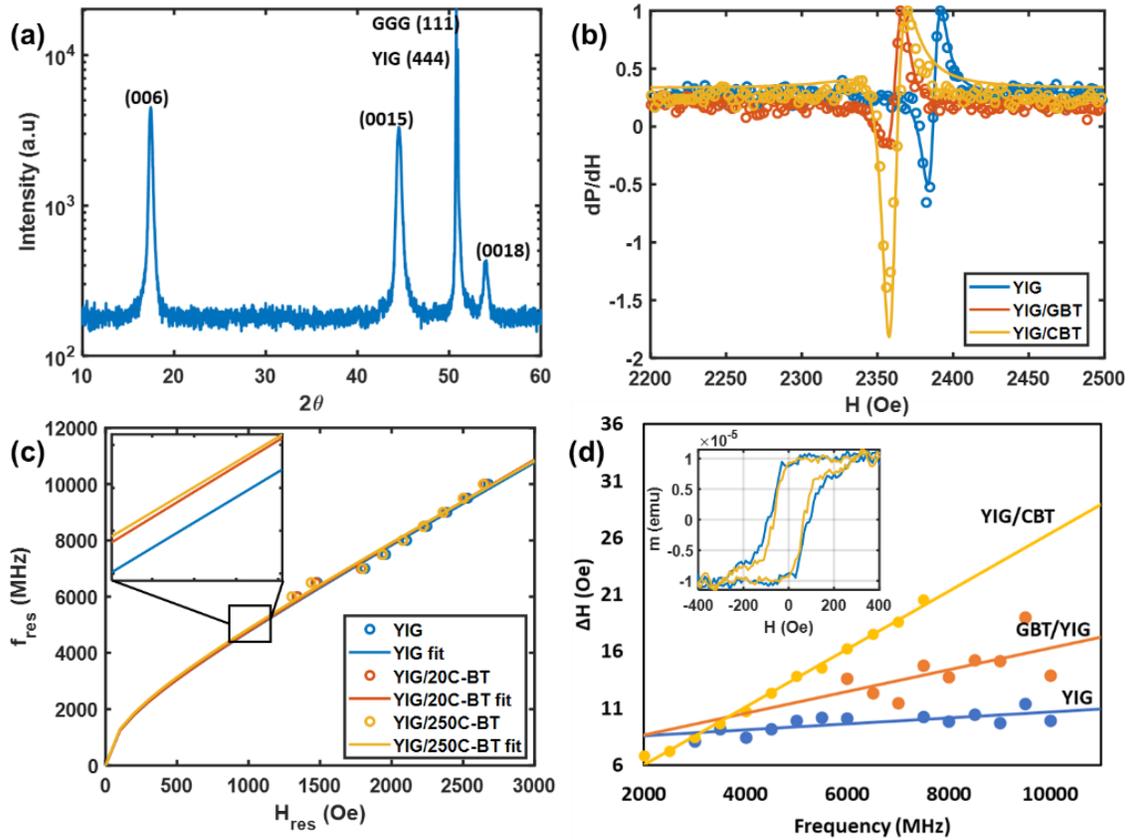

*Figure B6. a) XRD data for sputtered CBT grown on YIG showing strong c-axis oriented texture. b) FMR spectra of YIG/$Bi_2Te_3$ samples measured at 6 GHz showing enhancement in linewidth after deposition of GBT and CBT. c) FMR frequency versus resonance field plots and Kittel equation fitting of YIG, YIG/GBT and YIG/CBT samples. d) FMR linewidth versus frequency plots of YIG, YIG/GBT and YIG/CBT samples and linear fitting.*

**Table S3. Magnetic Properties of YIG/Bi$_2$Te$_3$ Samples.**

| Sample | $\frac{\gamma}{2\pi}$ | $\alpha$ | $4\pi M_s$ (kOe) |
|---|---|---|---|
| YIG | 2.81 | 0.000843 | 1.89 |
| GBT/YIG | 2.83 | 0.00283 | 1.89 |
| CBT/YIG | 2.84 | 0.00710 | 1.89 |